\documentclass[reprint,aps,prl,superscriptaddress,nofootinbib,floatfix]{revtex4-2}

\usepackage[T1]{fontenc}
\usepackage{amsmath,amssymb,bm,ulem}
\usepackage{graphicx}
\usepackage{xcolor}
\usepackage[colorlinks=true,citecolor=blue,linkcolor=blue,urlcolor=blue]{hyperref}

\newcommand{\athenak}{\textsc{AthenaK}}
\newcommand{\twopuncture}{\textsc{TwoPuncture}}
\newcommand{\SM}{Supplemental Material}
\newcommand{\figbox}[2][0.97\linewidth]{%
  \IfFileExists{#2}{\includegraphics[width=#1]{#2}}{%
    \fbox{\parbox[c][1.7in][c]{#1}{\centering Figure placeholder}}%
  }%
}

\begin{document}

\title{Trapping, Irregular Waveforms, and Efficient Radiation in Ultra-relativistic Black Hole Encounters}

\author{Hengrui Zhu}
\email{hengrui.zhu@princeton.edu}
\affiliation{Department of Physics, Princeton University, Princeton, New Jersey 08544, USA}
\affiliation{Princeton Gravity Initiative, Princeton University, Princeton, New Jersey 08544, USA}

\author{Frans Pretorius}
\affiliation{Department of Physics, Princeton University, Princeton, New Jersey 08544, USA}
\affiliation{Princeton Gravity Initiative, Princeton University, Princeton, New Jersey 08544, USA}

\author{James M. Stone}
\affiliation{School of Natural Sciences, Institute for Advanced Study, Princeton, New Jersey 08540, USA}

\begin{abstract}
We demonstrate that ultra-relativistic black hole encounters reveal a new regime of the two-body interaction in general relativity.
Evolving equal-mass, nonspinning black holes with initial center-of-mass Lorentz factors up to $\gamma\approx 5.1$ using numerical relativity, we find that the resulting waveforms defy the standard expectation of a post-Newtonian description followed by a smooth transition to a prompt Kerr ringdown. Instead, at nonzero impact parameter, the system can exhibit prolonged, highly irregular emission and significant horizon absorption, even without coalescence. We show these phenomena are driven by transient null trapping and repeated lensing of radiation in the binary interaction region.
Furthermore, our simulations indicate that over $65\%$ of the initial ADM energy can be radiated as gravitational waves at $\gamma\approx 5.1$, which is substantially larger than previously estimated by extrapolating from lower boost data.
\end{abstract}

\maketitle

\noindent {\bf \em Introduction.} 
Einstein's equations are strongly nonlinear, and for many years it was unclear how violently the merger of two black holes would behave once perturbative descriptions failed.
The breakthrough of numerical relativity showed that the astrophysical binary black hole problem is far more orderly than had been anticipated by some\footnote{Though not all; see e.g. \cite{Buonanno:2000ef}}: the inspiral is accurately captured by weak-field and post-Newtonian methods~\cite{Blanchet:2013haa}, the ringdown is controlled by perturbations of Kerr black holes~\cite{Teukolsky:1973ha,Berti:2009kk}, and the merger waveform smoothly interpolates between the two~\cite{Pretorius:2005gq,Campanelli:2005dd,Baker:2005vv,Centrella:2010mx,Sperhake:2014wpa}.
This empirical smoothness is itself suggestive.
In the astrophysical regime, the rest masses dominate the energy budget, the characteristic gravitational-wave frequencies remain comparatively low, and typically only a few percent of the total mass-energy is radiated.
The emitted radiation therefore backreacts only weakly on the spacetime and does not substantially reorganize the binary dynamics.

\begin{figure*}[t]
  \centering
  \figbox{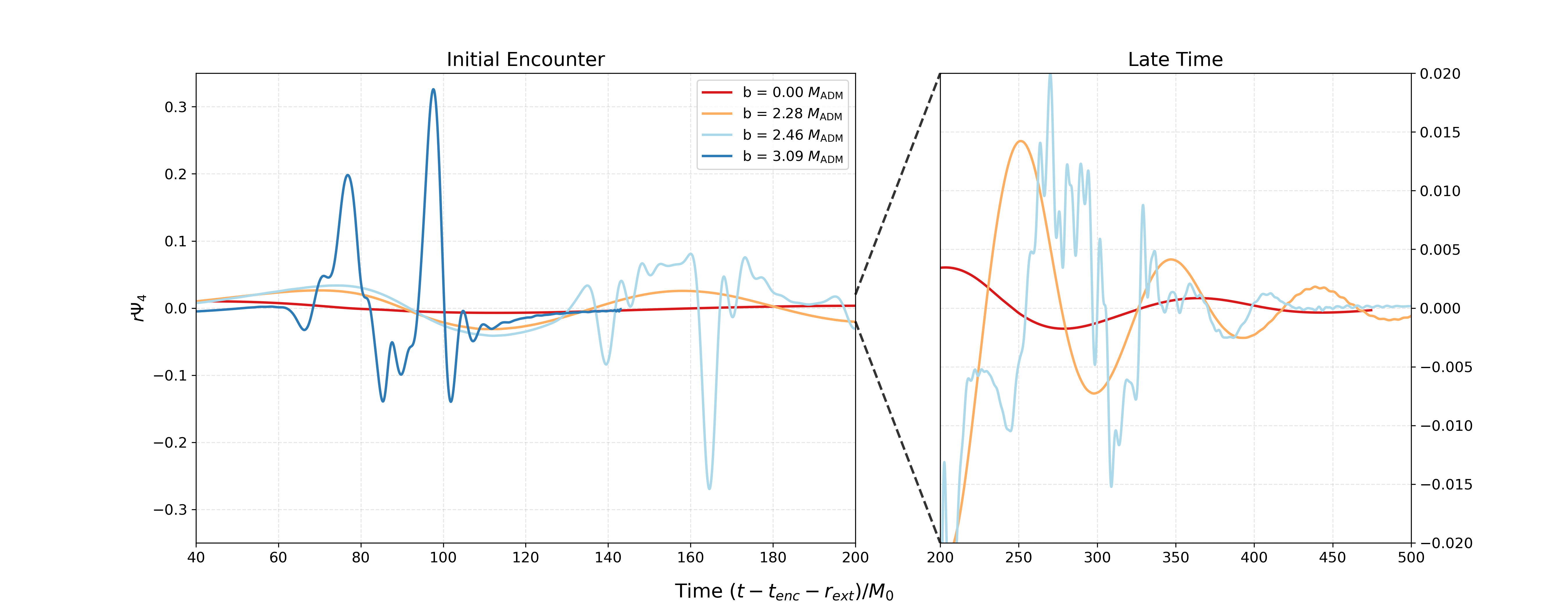}
  \caption{Representative $(\ell,m)=(2,2)$ mode of the extracted Weyl scalar $\Psi_4$ (real part) from four encounters at fixed $\gamma\approx 5.1$ and increasing impact parameter $b$: head-on collision, prompt merger, near-threshold zoom-whirl (no merger), and scattering (see the left panel of Fig.~\ref{fig:spectrum} for the corresponding coordinate trajectories). The imaginary part shares a similar amplitude envelope and morphology, except for the head-on case where it vanishes due to symmetry. The time axis is shifted by the extraction radius, $r_{\rm ext}=150 M_0$, and by an encounter time $t_{\rm enc}$, chosen to align the different impact-parameter cases. 
  The right panel uses an enlarged vertical scale to highlight the ringdown phase for the two cases where the black holes merged. (The very small amplitude high-frequency oscillations in the orange and light-blue curves at $t\gtrsim 400 M_0$ are due to spurious wave reflections at refinement boundaries, and converge away with increasing resolution). The progression across the different impact parameters illustrates the central phenomenology of this work.
  At small $b$, the signal resembles the familiar pattern of a burst followed by ringdown seen in a low-boost, near-head-on collision.
  Closer to the merger--scattering threshold, however, the waveform develops an extended and highly irregular wavetrain with multiple pulses.}
  \label{fig:waveform}
\end{figure*}

The distinction becomes sharper when viewed through the entropy bound supplied by Hawking's area theorem~\cite{Hawking:1971tu}.
For two equal-mass, initially nonspinning holes of irreducible mass $M_0$, one has $M_f\ge \sqrt{2}\,M_0$, so a rest-mass-dominated collision can radiate at most $\simeq 29\%$ of the total initial energy.
In realistic astrophysical mergers, the radiated fraction is much smaller, usually only a few percent~\cite{Centrella:2010mx,Sperhake:2014wpa}.
In the ultra-relativistic regime, however, the center-of-mass energy is $M_{\rm ADM}\simeq 2\gamma M_0$ while the irreducible masses remain fixed, so the same area theorem only implies
\begin{equation}
\frac{E_{\rm rad}}{M_{\rm ADM}}\le 1-\frac{1}{\sqrt{2}\gamma}~,
\end{equation}
which tends to unity as $\gamma\to\infty$.
In principle, then, nearly all of the energy can be converted into gravitational radiation without violating black hole thermodynamics \cite{Penrose1974,Sperhake:2008ga,Page:2022bem}.
This poses the ultra-relativistic encounter as an interesting regime of Einstein's equations, and though not directly applicable to astrophysical mergers, it is relevant to the physics of super-Planckian particle collisions~\cite{Dimopoulos:2001hw,Giddings:2001bu,Feng:2001ib,Choptuik:2009ww}.  

These considerations have motivated a broad literature on ultra-relativistic collisions, black hole formation, and strong-field scattering. Perturbative analytic studies of the axisymmetric head-on collision case estimate $\sim16\%$ radiative efficiency at infinite boost~\cite{DEath:1976bbo,DEath:1992hb,DEath:1992qu}. This was later shown to be consistent, to within $O(10\%)$, with extrapolations from full numerical evolutions of large but finite boost collisions\footnote{Similar results were found for ultra-relativistic collisions of ``soliton'' particle models~\cite{East:2012mb,Pretorius:2018lfb}, supporting the conjecture that in this limit ``matter does not matter''; this underlies the claims that pure classical gravity can describe the observable outcome of super-Planckian, small impact parameter particle interactions.}~\cite{Sperhake:2008ga,Healy:2015mla}. Extending to non-zero impact parameter collisions, analytic and semi-analytic studies emphasized the role of trapping and horizon formation~\cite{Eardley:2002re,Cardoso:2005jq}. Numerical relativity studies of finite-impact-parameter encounters with boosts up to $\gamma\sim 2.9$ revealed significantly higher radiative efficiency than head-on collisions, zoom-whirl behavior near the prompt-merger threshold, and showed that even without coalescence in the scattering regime a substantial fraction of the center-of-momentum energy could be absorbed by the individual black holes~\cite{Shibata:2008rq, Sperhake:2009jz, Sperhake:2012yc}. Zoom-whirl dynamics was also found in the extreme mass ratio limit using self-force calculations~\cite{Gundlach:2012aj}, which also supported the conjecture that near-perfect efficiency of kinetic to gravitational wave energy conversion occurs at threshold. More recent work~\cite{Healy2025} has broadened the search for maximal radiative efficiency and remnant spin, but the explored finite-impact-parameter encounters were only mildly relativistic compared with the asymptotic regime of interest.

Pushing to higher Lorentz factors using numerical relativity has proven difficult. 
The challenge is not only the severe length contraction of the fields, but also the gauge response to singular puncture data with extreme momentum. 
In earlier evolutions of high-boost Bowen--York initial data, the lack of clean convergence was often discussed primarily in terms of high frequency, so-called ``junk'' radiation~\cite{Healy:2015mla}. 
Our calculations suggest that this is not the ultimate issue: gauge dynamics themselves become a leading obstruction, both at the initial time where the isotropic slicing collapses to trumpets~\cite{Hannam:2008sg,Slinker:2018ujj}, and near the initial encounter through the generation of strong secondary gauge pulses. The main technical advance of this work is a new telegrapher-type lapse driver, described elsewhere in detail~\cite{ZhuTruongInPrep}, that remains stable in this ultra-relativistic regime.

The central physical picture that emerges is that small-impact-parameter encounters create strongly self-gravitating, thin gravitational-wave packets that are initially ``pancake'' shaped and then evolve and interact in a region of spacetime featuring (transient) null trapping. 
The longitudinal scale of the initial wave packets is set by the Lorentz-contracted irreducible mass of each black hole $M_0/\gamma$, while the transverse scale and size of the interaction region are of order the net gravitational mass of the spacetime $2M_0\gamma$ (and therefore the null trapping referred to here is {\em not} simply associated with the light-rings of the individual black holes). 
The null trapping, and interaction of the lensed wave packets with the two black holes that are likewise temporarily ``trapped'', causes several distinct collisions of the wave fronts with each other and the individual black holes. 
The result is a highly irregular waveform that is not well-captured by the simple quadrupole emission picture, together with substantial energy absorption and corresponding horizon growth by the individual black holes. In this Letter, we present the numerical results that led to this qualitative picture.

Furthermore, we show that, likely due to the dynamics just described, for boosts above $\gamma \sim 3$ the impact parameter resulting in maximum radiation efficiency starts to deviate from that corresponding to the critical threshold separating merger from scattering. This invalidates earlier assumptions used to extrapolate finite boost data to the infinite $\gamma$ limit~\cite{Sperhake:2012yc}, significantly underestimating the radiated fraction of over $65\%$ that we find here at $\gamma\approx 5.1$. Our results also show $\gamma\approx 5.1$ is not yet high enough to get a clear picture of the asymptotic limit.

\begin{figure*}[t]
  \centering
  \figbox{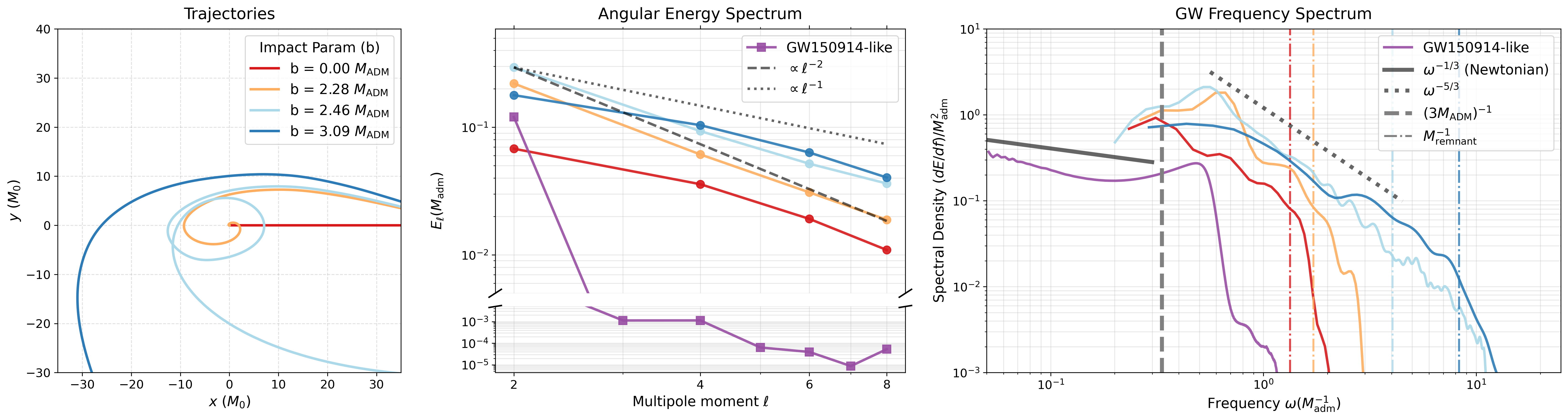}
  \caption{\textit{Left:} coordinate trajectory of one of the black holes from each of the four cases shown in Fig.~\ref{fig:waveform}, illustrating the progression from head-on collision to prompt merger, near-threshold zoom-whirl, and scattering.
  \textit{Middle:} angular-mode energy spectrum of the emitted gravitational radiation.
  The power falls off with multipole number much more slowly than in astrophysical quasi-circular binary mergers, an example of which is shown with the purple line~\cite{LIGOScientific:2016aoc,Radice:2025djo}, indicating that the radiation is distributed across a broad hierarchy of angular scales rather than being dominated by only the lowest modes.
  \textit{Right:} frequency-domain energy spectrum of the emitted gravitational waves.
  The spectrum is broad and retains substantial high-frequency support with a spectral slope that resembles Kolmogorov turbulence in fluid dynamics, with an $\omega^{-5/3}$ scaling~\cite{Kolmogorov_1968}, distinct from the $\omega^{-1/3}$ expectation of a Newtonian quasi-circular inspiral.
  Vertical dashed lines indicate key characteristic frequency scales.
  The gray dashed line marks the spacetime's dominant low-frequency scale, corresponding to the trapping region's light-crossing time.
  Colored dashed lines denote the size of the end state black hole horizons: the single remnant for prompt mergers, and the enlarged individual horizons following strong absorption in scattering encounters.
  The frequency corresponding to the Lorentz-contracted curvature scale (roughly $50\,M_{\rm ADM}^{-1}$, see the first panel of Fig.~\ref{fig:superpoynting}) lies above the plotted range.
  }
  \label{fig:spectrum}
\end{figure*}

\noindent {\bf \em Methods.} 
The calculations are performed with \athenak~\cite{stone2024,Zhu2025,Fields2025}, a performance-portable extension of the Athena++/GR-Athena++ framework~\cite{Stone:2020aow,Daszuta:2021ecf} designed for modern heterogeneous supercomputers.
The Einstein equations are evolved in the Z4c formulation, whose constraint-propagation and damping properties make it well suited to puncture evolutions~\cite{Bona:2003fj,Bernuzzi:2009ex,Weyhausen:2011cg,Hilditch:2012fp}.
\athenak\ uses block-structured octree adaptive mesh refinement together with Kokkos-based portability, allowing the same implementation to run efficiently on large GPU systems.

Our baseline gauge conditions are an extension of those of the moving-puncture paradigm, namely 1+log slicing for the lapse together with a Gamma-driver shift, the same combination that underpinned the 2005--2006 numerical relativity breakthroughs for BSSN codes~\cite{Campanelli:2005dd,Baker:2005vv}.
In strongly dynamical puncture evolutions, however, Bona-Masso-type slicing conditions are known to develop steep gauge features and even shock-like pathologies~\cite{Alcubierre:1996su,Alcubierre:2004hr,Baumgarte:2022ecu}.
A slow-start modification of the lapse source, recently proposed in Ref.~\cite{Etienne:2024ncu}, can attenuate the initial gauge transition, but in the present problem it does not remove the later emergent pulses generated when two highly boosted punctures pass through one another's strong-field region\footnote{A shock-avoiding slicing condition was proposed in Ref.~\cite{Alcubierre:1996su}. However, it is known to create regions with negative lapse, complicating physical interpretation despite seeming numerical stability~\cite{Baumgarte:2022ecu}.}.
The telegrapher gauge used here supplements the lapse evolution with an auxiliary field so that the principal part becomes a damped telegrapher equation: gauge distortions propagate away and decay rather than steepen catastrophically.
We do not reproduce the full equations here; the formulation, characteristic structure, and calibration will appear in an upcoming paper~\cite{ZhuTruongInPrep}.

Initial data are constructed from equal-mass Bowen--York punctures using the \twopuncture\ code~\cite{Bowen:1980yu,Brandt:1997tf,Ansorg:2004ds}.
Conformal flatness inevitably injects spurious gravitational radiation into boosted puncture data (this is the so-called ``junk'' radiation), but for the present purposes Bowen--York data remain practical and informative.
In the high-boost regime, the dominant effect of this junk burst is to renormalize the relation between the puncture momentum and the physical boost.
For the boosts considered here this is an $\mathcal{O}(10\%)$ effect, which is not large enough to preclude the use of Bowen--York initial data for such large boosts\footnote{This is in contrast to the sensitivity to initial-data artifacts for near-extremal spin problems, where even sub-percent level spurious absorption can push the resultant physical spins too far from the near-extremal regime to be of interest~\cite{Lovelace:2008tw,Liu:2009al,Ruchlin:2014zva}.}.
We therefore calibrate the physical Lorentz factor by fitting the ADM mass as a function of separation, estimating the junk content and the binding energy (for extracting the effective boost factor at infinite separation), and subtracting their contributions from the energy budget.
The details of this procedure, together with convergence tests and the asymptotic scaling of the Bowen--York boost junk, are summarized in the \SM.

Note however that we have {\em not} calibrated the initial coordinate impact parameter $b$, measured from the puncture coordinates on the initial slice, to the asymptotic impact parameter one would like in the scattering-theory sense. Therefore $b$ should be considered as a simulation label that only loosely corresponds to the physical impact parameter. 
Lastly, we use geometrized units, with $G=c=1$, and set the irreducible mass of the initial black holes, $M_0$, to be unity.

\noindent {\bf \em Waveform morphology.} 
In ultra-relativistic black hole encounters, the gravitational waveform is qualitatively unlike the familiar signal from astrophysical mergers.
Fig.~\ref{fig:waveform} shows representative $(\ell,m)=(2,2)$ waveforms from four encounter regimes at a boost of $\gamma=5.1$.
Rather than a short burst followed by prompt settling into Kerr ringdown, the non-prompt-merger signals, namely the two larger-impact-parameter cases, develop prolonged, irregular wavetrains with multiple delayed pulses.
This behavior is strongest near the prompt-merger/scattering threshold, where the system spends the longest time in the nonlinear near zone.

\begin{figure*}[t]
  \centering
  \figbox{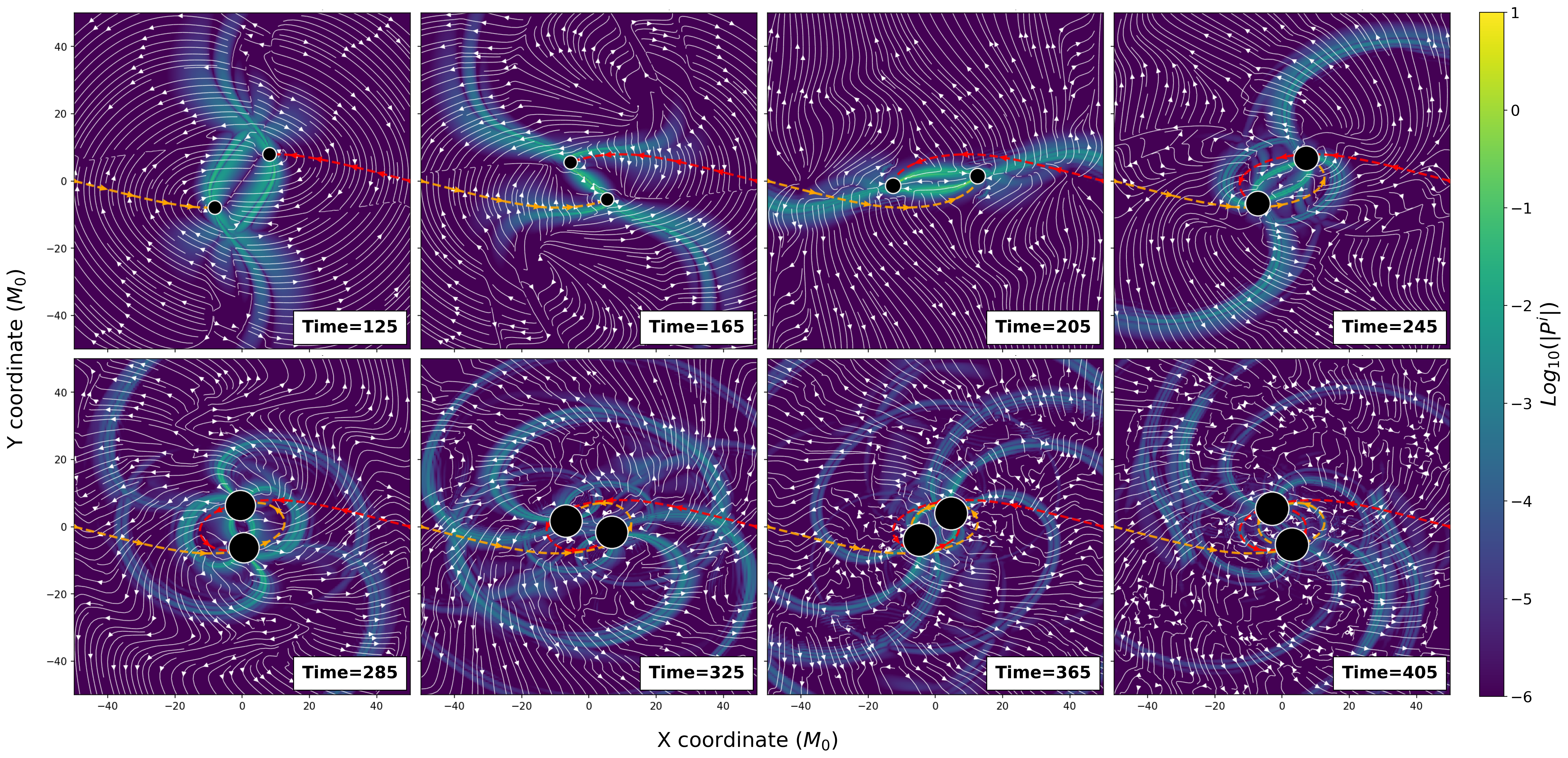}
  \caption{Snapshots of the Bel--Robinson super-Poynting flux $|\bm{P}|$ on the orbital plane at different times (in units of $M_0$) for a near-threshold encounter with $b=2.46 M_{\rm ADM}$. (See Fig.~\ref{fig:waveform} for the corresponding waveform at large radius.)
  The dashed curves mark the past trajectories of the black holes (circles), with the circle sizes scaled to represent the instantaneous irreducible masses of the black holes as measured from their apparent horizons~\cite{Ashtekar:2004cn,Thornburg:2003sf}.
  Horizon properties as a function of evolution time for this encounter are detailed in Fig.~\ref{fig:absorption}.
  During the early approach, each boosted hole carries a strongly Lorentz-contracted near-zone curvature pattern.
  The ``curvy'' nature of the wave fronts in the initial frame is in part a gauge effect.
  Following the initial close encounter, this curvature flux is focused, sheared, and repeatedly lensed within the interaction region, eventually producing the intricate flux pattern that corresponds to the delayed pulses seen in Fig.~\ref{fig:waveform}.
  At late times, both horizons have grown substantially through absorption, and the black holes fly apart with a mildly relativistic speed of $\sim 0.13c$.}
  \label{fig:superpoynting}
\end{figure*}

\begin{figure}[t]
  \centering
  \figbox{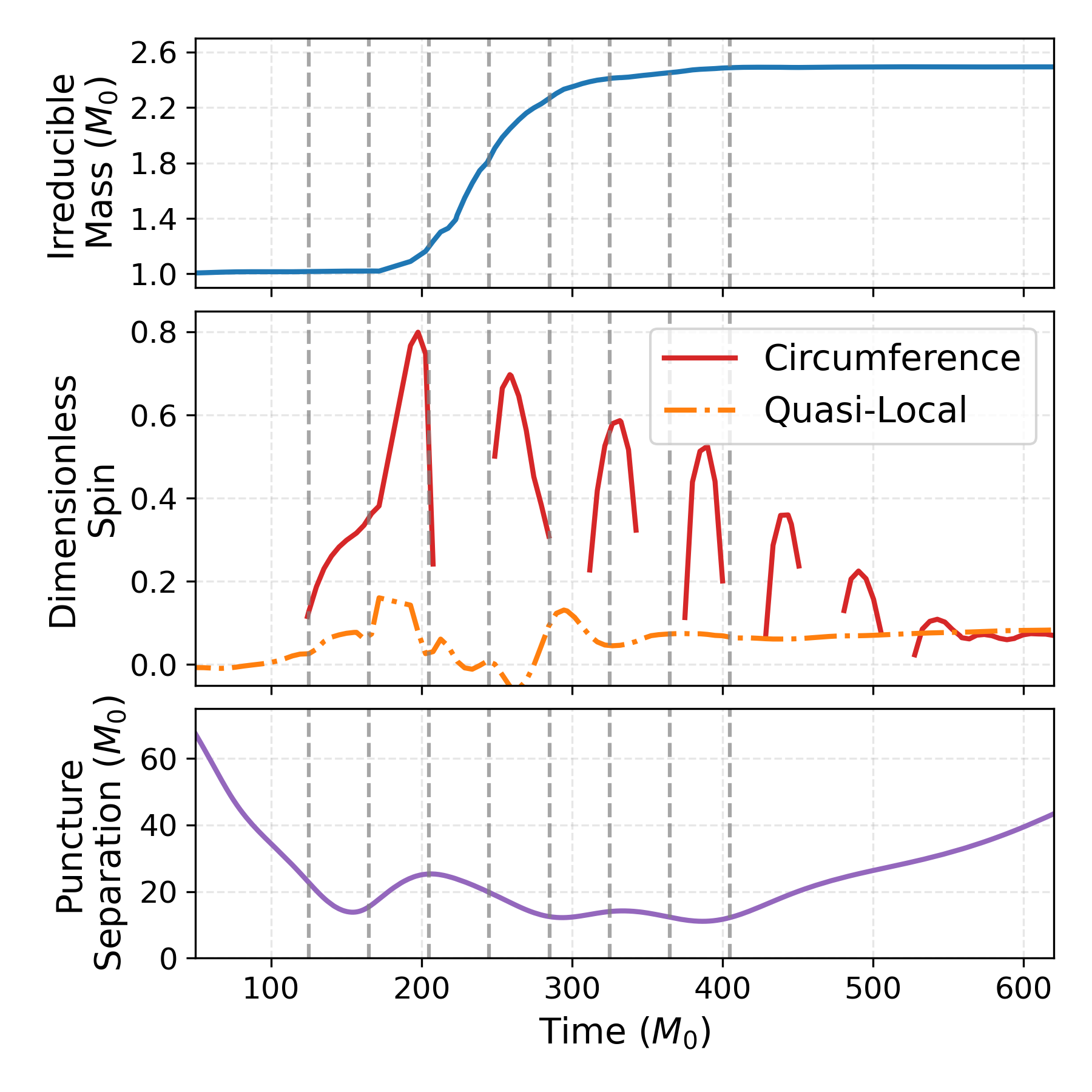}
  \caption{\textit{Top:} Irreducible mass for one of the black holes for the near-threshold, non-merger encounter shown in Fig.~\ref{fig:superpoynting}.
  The vertical dashed lines represent the times at which the eight frames in Fig.~\ref{fig:superpoynting} were chosen.
  Despite the lack of merger, the irreducible mass grows by a factor of roughly 2.5 through the accretion of gravitational radiation.
  \textit{Middle:} Dimensionless spin inferred through the quasi-local dynamical-horizon measure (dashed orange) and equatorial circumference (solid red) measures---see Eqs.~(\ref{spin_local}) and (\ref{spin_ce}), respectively, in the Supplemental Material.
  During the encounter, the black holes are significantly distorted, yielding large oscillations in the spin inference, in particular from the circumference estimate (for which the Kerr-based inversion can become ill-defined, producing gaps in the red curve).
  At late times, the two estimates asymptote to the same value.
  \textit{Bottom:} Coordinate separation between the two black holes as a function of time.}
  \label{fig:absorption}
\end{figure}

\begin{figure}[t]
  \centering
  \figbox{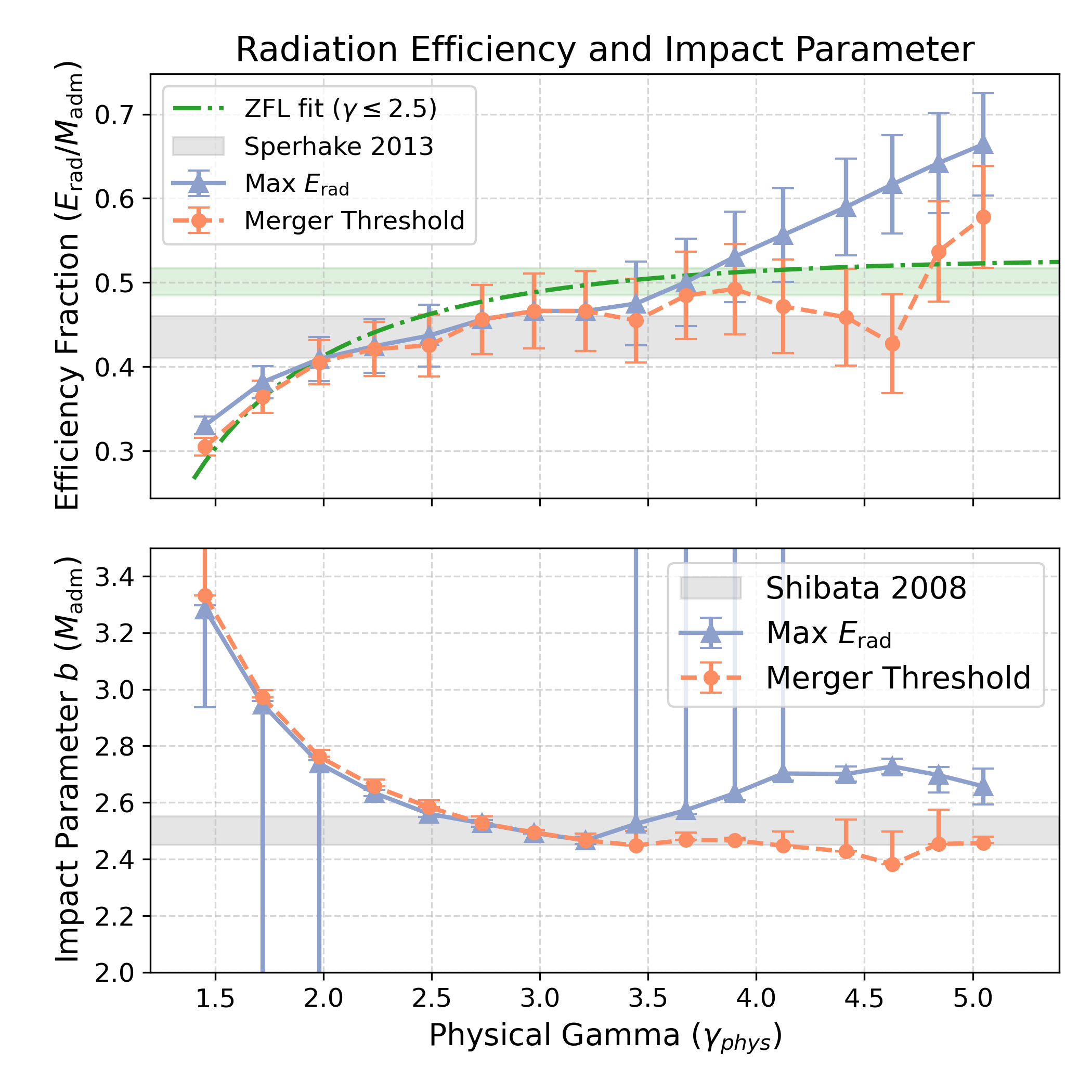}
  \caption{\textit{Top:} Radiative efficiency, defined as the fraction of the physically relevant ADM mass converted into escaping gravitational radiation ($E_{\rm rad}/M_{\rm ADM}$), for the impact parameter with maximal emitted energy (purple) and the critical impact parameter separating merger from scattering (orange).
  Error bars indicate conservative systematic uncertainties associated with the residual junk-radiation subtraction and do not reflect the sampling uncertainty in the impact parameter (though as implied by the panel below, for larger $\gamma$ the gap between the orange and purple curves is well resolved even after accounting for the latter.)
  The gray shaded region is from Ref.~\cite{Sperhake:2012yc}, where extrapolation from $\gamma\leq 2.5$ suggested that roughly 50\% of the ADM mass is converted into gravitational radiation in the infinite-$\gamma$ limit.
  The green dashed curve shows the extrapolation from our maximal radiation data up to $\gamma=2.5$ using the functional form of the zero-frequency limit (ZFL)~\cite{Smarr:1977fy,Berti:2010ce,Sperhake:2008ga,Sperhake:2012yc}, with the green shaded region showing the asymptotic value and fitting uncertainty (note that the ZFL-based expression is not monotonic in $\gamma$, and the asymptotic value is slightly less than at $\gamma\approx 5.1$).
  The low boost extrapolation diverges significantly from our high-$\gamma$ data.
  \textit{Bottom:} Impact parameters $b$ that maximize the radiated energy (purple) and separate scattering from merger (orange) as a function of $\gamma$.
  The gray shaded region shows extrapolation from lower boost data in Ref.~\cite{Shibata:2008rq}. 
  Asymmetric error bars indicate the sampling resolution in impact parameter.}
  \label{fig:radiation_efficiency}
\end{figure}

The spectrum also highlights the distinction from astrophysical mergers.
Fig.~\ref{fig:spectrum} shows both the orbital morphology and the radiative spectra for the different encounter regimes. Compared with astrophysical binary mergers, the spectra feature more power at high frequencies, and display a distinctive spectral slope similar to that of the Kolmogorov spectrum in fluid turbulence~\cite{Kolmogorov_1968}.
The former property is natural in the ultra-relativistic regime: Lorentz contraction introduces a new short length scale into the problem, while the effective trapping scale, set roughly by the boosted center-of-mass energy, grows with the total ADM mass.
The regime is therefore more favorable to nonlinear self-interaction compared to astrophysical mergers: more energy is available to radiate, and the emitted waves have a wavelength component that is small compared with the scale of the binary near zone, allowing for repeated strong lensing that forces self-interaction.

A useful heuristic is to compare the initial longitudinal packet scale, $M_0/\gamma$, with the scale over which we expect transient null trapping, following a near threshold impact parameter encounter.
A natural estimate for the latter scale is obtained by assuming that the boosted center-of-mass energy sets a characteristic radius $R_{\rm trap}\sim 3M_{\rm tot}$, giving a transient binary ``light-ring region'' of size $\sim 6\gamma M_0$ for an equal mass encounter.
Thus, the ratio of the longest to shortest scales in the problem grows like $\gamma^2$.
Radiation produced within a region of radius $R_{\rm trap}$ does not promptly radiate away.
Depending on where on the wavefront one looks, the radiation can be flung outward, turned back toward the source black hole, sheared between the two black holes, or captured by the companion.

\noindent {\bf \em Strong-Field Geometrodynamics.} 
To make the trapping picture more concrete, we analyze a local curvature diagnostic based on the electric and magnetic parts of the Weyl tensor, $E_{ij}$ and $B_{ij}$, namely the Bel-Robinson super-Poynting vector $\bm{P}^i\equiv \epsilon^{ijk}E_{jm}B^{m}{}_k$~\cite{Owen:2010vw,Nichols:2011pu,Zhang:2012pe,Zhang:2012ky,Nichols:2012tq,Bel:1958,Robinson:1997,Maartens:1997fg}, which is the analog of the familiar Poynting flux in electromagnetism\footnote{For a recently proposed formalism that establishes an even stricter parallel between curvature dynamics and classical electrodynamics, see Ref.~\cite{Boyeneni:2025tsx}}.
The amplitude of the super-Poynting flux is related to the spin-2 Weyl scalars in a transverse frame~\cite{Zhang:2012pe}:
\begin{equation}
    |\bm{P}| \propto \frac{1}{2}
    \left\vert|\Psi_4|^2 - |\Psi_0|^2\right\vert~.
\end{equation}
This does not replace wave extraction at large radius using the Newman--Penrose scalars, but is useful for diagnosing how curvature is transported and redirected in the nonlinear near zone.
For an isolated boosted black hole, the super-Poynting flux is the convected curvature of the black hole, rather than an actual radiative flux.
Following the encounter, a significant fraction of this convected curvature is converted into a strong, initially plane-fronted gravitational-wave flux, analogous to Bremsstrahlung radiation in electromagnetism.

Fig.~\ref{fig:superpoynting} shows snapshots of the super-Poynting flux $\bm{P}^i$ on the orbital plane at several times for the run closest to the critical impact parameter that separates prompt merger from scattering (with $\gamma\approx 5.1$).
The two black holes complete two orbits, more than doubling their irreducible mass through absorption (see Fig.~\ref{fig:absorption}), before flying apart to infinity with a mildly relativistic speed of $\sim 13\%~ c$.
These snapshots qualitatively explain the highly irregular waveform shown in Fig.~\ref{fig:waveform}.
During approach, the dominant curvature flux is advected with the individual holes, with little outgoing radiation produced.
Following the initial close passage, the wave fronts are violently sheared, redirected, and folded around the binary.
The outer edges of the wave fronts peel off to infinity, while the inner parts are temporarily trapped in the interaction region, with some of the latter absorbed by the black holes.
The \SM\ includes accompanying volume renderings and null-geodesic visualizations, supplementing the discussion here.

\noindent {\bf \em Radiation efficiency.} 
The complex near-zone dynamics fundamentally alter how kinetic energy is partitioned into escaping radiation versus horizon absorption.
Because gravitational waves generated deep inside the transient trapping region do not simply escape but are repeatedly lensed and partially absorbed by the horizons, the radiative efficiency does not scale trivially with the impact parameter and initial Lorentz boost.

Fig.~\ref{fig:radiation_efficiency} summarizes the radiative efficiency as a function of the initial Lorentz boost, which we define as the fraction of the physically relevant ADM mass converted into escaping gravitational waves, $E_{\rm rad}/M_{\rm ADM}$ (with spurious initial data junk excluded from the energy budget).
We estimate numerical truncation error to be strictly subdominant to the conservative systematic uncertainties associated with this junk subtraction, which are reflected in the error bars.

Crucially, the top panel of Fig.~\ref{fig:radiation_efficiency} demonstrates that the maximum radiative efficiency strongly diverges from previously conjectured extrapolations~\cite{Sperhake:2012yc}.
Ref.~\cite{Sperhake:2012yc}, relying on data up to $\gamma \le 2.5$, suggested that roughly $50\%$ of the kinetic energy is radiated in the infinite-boost limit, with the remainder absorbed.
This extrapolation relied on two key assumptions: first, that the underlying physical mechanism for absorption was sufficiently strong focusing of gravitational waves {\em toward} the individual black holes so that the accretion rate would scale with $\gamma$, despite the fact that the initial absorption cross section (proportional to the horizon area) does not; and second, that the maximal radiation happens at the threshold of merger.
Our results demonstrate that both of these assumptions break down. 
Most notably, we have explicit numerical solutions that feature over $65\%$ of the initial ADM energy radiated as gravitational waves (top panel of Fig.~\ref{fig:radiation_efficiency}), well above the extrapolated limit.
Moreover, the maximum radiation branch clearly separates from the merger threshold at these higher boosts (bottom panel of Fig.~\ref{fig:radiation_efficiency}). Regarding the accretion mechanism, by comparing the irreducible mass growth curve shown in Fig.~\ref{fig:absorption} to the Poynting flux plots in Fig.~\ref{fig:superpoynting}, it is more suggestive that accretion occurs when the black holes pass through wakes of the strongly lensed gravitational wave packets (i.e. there is very little growth in area until the third panel of Fig.~\ref{fig:superpoynting}, which is the first time each black hole traverses a lensed wake). Together, these results and our interpretation imply that $\gamma\approx 5.1$ is still not high enough to allow a well-justified extrapolation to $\gamma=\infty$.

\noindent {\bf \em Discussion.} 
The simulations reported here reveal a regime of binary dynamics that is qualitatively different from the one most familiar from gravitational-wave astronomy.
This distinctive regime is characterized by more prolonged and irregular waveforms, broad high-frequency spectra, and repeated wave--black-hole and wave--wave collisions inside the binary near zone.
In such a regime, the emitted radiation does not immediately leave the system once generated; it repeatedly lenses through the binary and is slowly partitioned between escape to infinity and horizon absorption on the orbital timescale.

Several steps for future work are evident.
First, our results firmly establish that current numerical solutions are not yet in a regime where a reliable extrapolation to $\gamma\to\infty$ can be made.
Because horizon absorption seems to depend intimately on the complex multi-pass lensing of curvature, establishing the true asymptotic scaling will require understanding how the number of near-zone interactions scales with the initial boost.
This, together with tracking how the impact parameter of maximum radiation efficiency separates from the scattering threshold as a function of impact parameter, will require simulations at even higher Lorentz factors and larger initial separations.
Second, for trajectories that result in prompt mergers, the spin of the remnant in the ultra-relativistic limit remains an important open question.
Recent lower-boost parameter surveys indicate that large remnant spins are possible in high-speed collisions~\cite{Healy2025}; determining the corresponding limit in the deeply ultra-relativistic regime will be a priority for future work.

Finally, our results suggest that the apparent ``only mildly nonlinear'' nature of the observable emission from astrophysical binary mergers is not a generic property of the two-body interaction in general relativity. 
In the astrophysical context, the radiated energy fraction is small, essentially driven by the quadrupolar acceleration of the black holes, and the characteristic wavelengths are comparable to or larger than the orbital scale, so the binary remains in the near zone of the emitted radiation.
In the ultra-relativistic regime the situation is very different: 
Lorentz contraction introduces a short-wavelength component into the radiation that can be (arbitrarily) small compared to the orbital scale, and more than $65\%$ of the center-of-momentum energy can be converted into gravitational radiation.
This implies that the interaction of the strongly self-gravitating radiation is more important than any subsequent quadrupole-like emission from the two black holes in describing the resulting waveform. 
Ultra-relativistic encounters therefore peel back the smooth facade of the two-body problem, revealing a fully nonlinear, strongly self-interacting sector of general relativity that remains hidden in astrophysical mergers.

\noindent {\bf \em Acknowledgments.} We thank David Radice, Sebastiano Bernuzzi, William East, Abhishek Hegde, Gautam Satishchandran, Nils P. Siemonsen, and Elias Most for insightful discussions.
H.Z. also thanks Zachariah B. Etienne for valuable discussions on gauge conditions and horizon finders, Robert Owen for discussions of the vortex-tendex formalism, and Haiyang Wang for discussions regarding 3D visualization.
Finally, we thank our INCITE coordinator Kyle Felker and Michael Buehlmann for their extensive assistance with running on the Aurora Exascale Supercomputer hosted at ALCF.
An award of computer time was provided by the U.S. Department of Energy’s (DOE) Innovative and Novel Computational Impact on Theory and Experiment (INCITE) Program, under project \textbf{RadBlackHoleAcc} and \textbf{CompactBinaryMerger}.
This research used resources from the Argonne Leadership Computing Facility, a U.S. DOE Office of Science user facility at Argonne National Laboratory, which is supported by the Office of Science of the U.S. DOE under Contract No. DE-AC02-06CH11357.
This work used Delta/DeltaAI at the National Center for Supercomputing Applications through allocation PHY240301 from the Advanced Cyberinfrastructure Coordination Ecosystem: Services \& Support (ACCESS) program, which is supported by U.S. National Science Foundation grants \#2138259, \#2138286, \#2138307, \#2137603, and \#2138296.
Early test simulations for this Letter were also performed on computational resources managed and supported by Princeton Research Computing, a consortium of groups including the Princeton Institute for Computational Science and Engineering (PICSciE) and Research Computing at Princeton University. FP acknowledges support from the NSF through the grants PHY-220728 and PHY-2512075.
\appendix
\section*{Supplemental Material}

\begin{figure*}[t]
  \centering
  \includegraphics[width=1\textwidth]{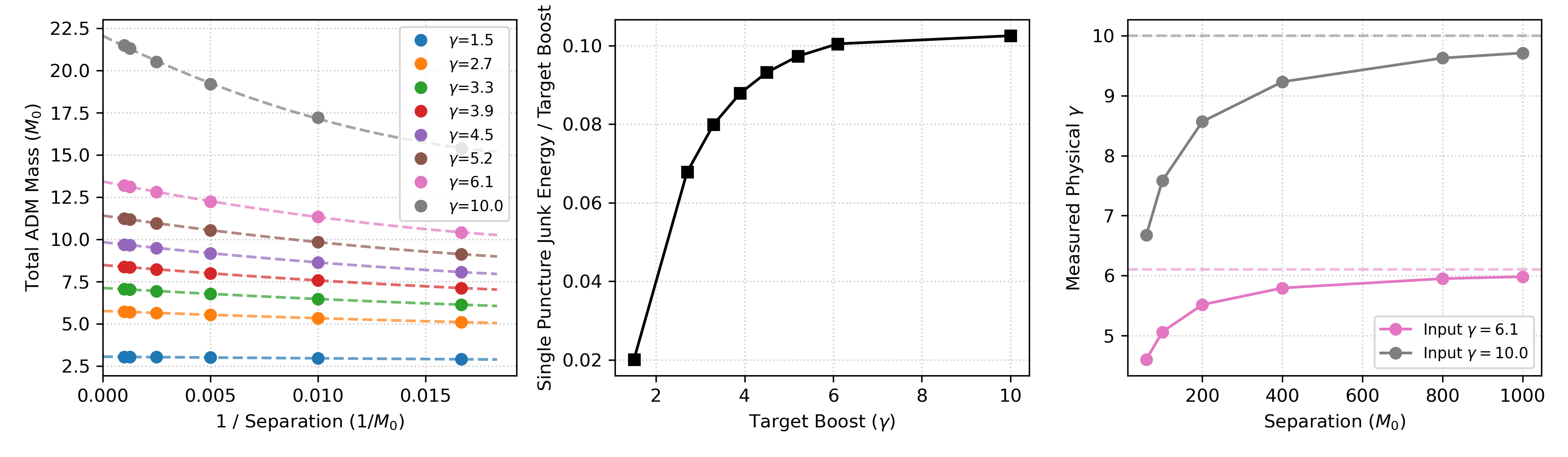}
  \caption{\textit{Left:} ADM mass as a function of inverse separation for a range of input puncture Lorentz factors. The dashed lines represent fits with the functional form in Eq.~\ref{eq:fit}. \textit{Middle}: Junk radiation as a function of the input boost (through the puncture momentum) \textit{Right}: The extracted physical boost for fixed input puncture momentum as a function of separation.}
  \label{fig:id}
\end{figure*}

\section{Bowen--York Initial Data with Ultra-relativistic Boost}
In previous works, Bowen--York initial data were deemed problematic at ultra-relativistic momentum due to the conformally flat assumption~\cite{Ruchlin:2014zva, Healy:2015mla}, similar to the case of high spin. 
Here, we show that this is not the case. 
For a single boosted puncture under the Bowen--York ansatz, one can solve the Hamiltonian constraint iteratively while keeping the irreducible mass of the black hole, as measured from the other asymptotic end, fixed to unity by adjusting the bare mass parameter.
The momentum of the horizon can be measured on the initial slice, which is maximal and conformally flat, through the quasi-local formula proposed in Ref.~\cite{Krishnan:2007pu}. 
The quasi-local momentum measured at the apparent horizon agrees very well with the Bowen--York momentum.
However, we find numerically that the total ADM mass of the spacetime measured at spatial infinity yield a value that is consistently larger than $\gamma M_0$ and increases monotonically with $\gamma$. 
The excess amount is due to the presence of junk radiation, which accounts for roughly 10\% of the ADM mass at a boost of $\gamma=10$. 
Subsequent evolution of this data shows that the horizon does not absorb the junk radiation efficiently (see e.g. Fig.~4 from the main text), as the irreducible mass of the black hole changes less than 1\% even at a boost of $\gamma\approx5.1$. 

The interpretation becomes a bit more complicated in the presence of another black hole. 
We find numerically that at large separation, the excess ADM energy of a binary, inferred using this technique, asymptotes to the same value as the single black hole case. 
However, at finite separation, a negative term, analogous to the Newtonian binding energy, also contributes to the energy budget. 
Making the assumption that the junk radiation is roughly constant at finite but large separation, one can distinguish the contribution of the binding energy and junk radiation by fitting the ADM mass as a function of separation at each fixed puncture momentum. 
The estimate for the binding energy is important in that this term directly affects the inferred physical $\gamma$ at infinite separation. 
In our fitting formula, we also include the leading-order nonlinear term, which scales as the inverse square of the separation. 
The whole formula reads:
\begin{equation}\label{eq:fit}
    M_{\rm ADM}^{\rm total}(d) = 2\gamma_0M_0 + E_{\rm junk} + \frac{C_1}{d} + \frac{C_2}{d^2} + \mathcal{O}(d^{-3})~,
\end{equation}
where $\gamma_0 = \sqrt{1+\left(\frac{P}{M_0}\right)^2}$ with $P$ being the amplitude for the Bowen--York momentum, $E_{\rm junk}$ is the amplitude of junk radiation, and $C_1$ and $C_2$ are coefficients for the binding-energy and leading-order nonlinear contributions. 
In the main text, we use the junk subtracted ADM mass $M_{\rm ADM} = M_{\rm ADM}^{\rm total} - E_{\rm junk}$. 
Then, we can estimate the effective physical $\gamma$ at infinite separation:
\begin{equation}
    \gamma = \frac{M_{\rm ADM}^{\rm total} - E_{\rm junk}}{2M_0}
\end{equation}

We find numerically that the ADM mass of the ultra-relativistic binary spacetime is independent of impact parameter.
The initial data diagnostics are further summarized in Fig.~\ref{fig:id}.
For the runs presented in the main text, we used an initial separation of $d=100M_0$ along the $x$ axis, and $\gamma_0=6.1$, which yields a physical Lorentz factor of $\gamma\approx 5.1$. 

\section{Radiation efficiency as a function of impact parameter}

\begin{figure*}[t]
  \centering
  \includegraphics[width=0.8\textwidth]{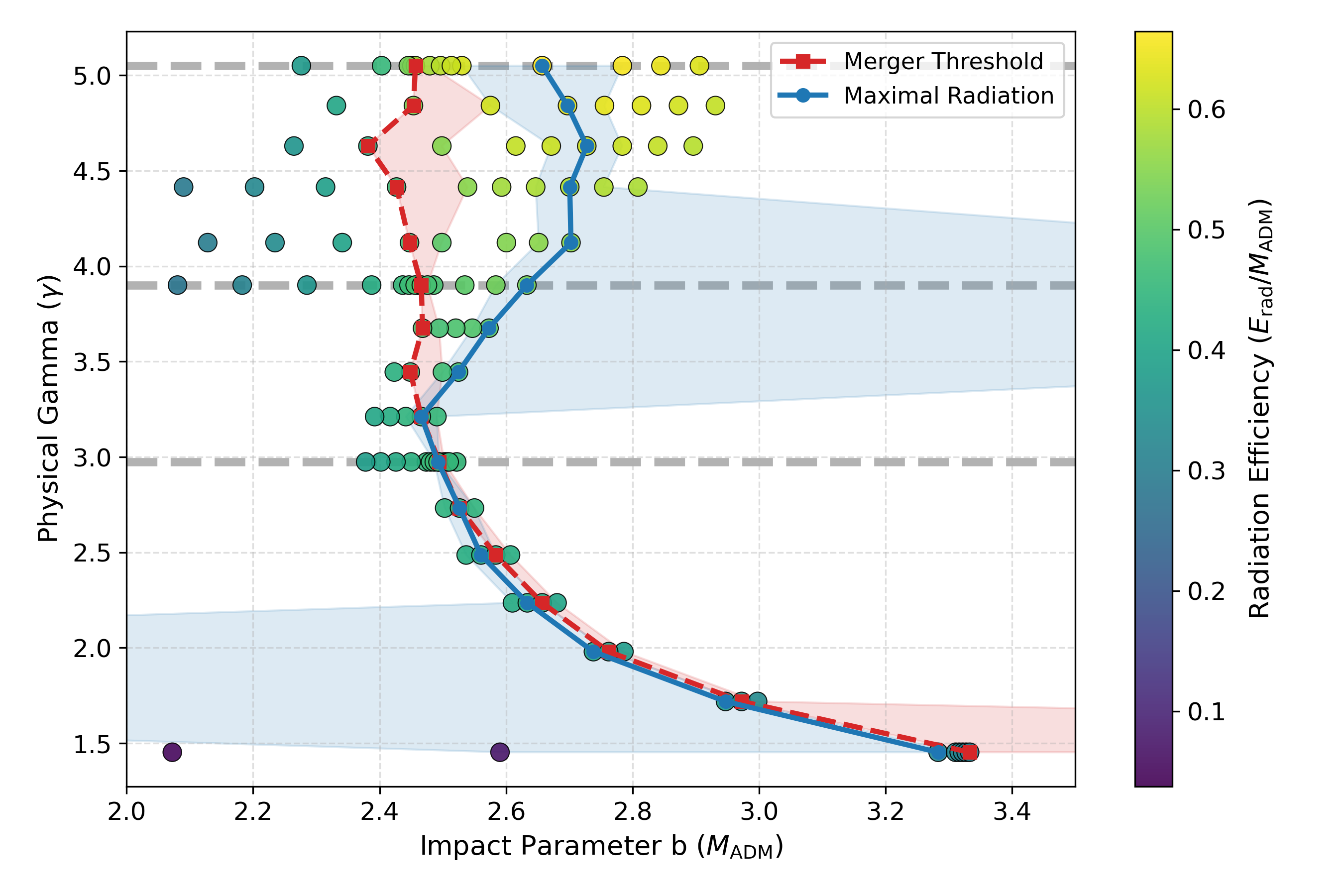}
  \caption{Radiation efficiency as a function of impact parameter and Lorentz factor. The red dashed line shows the merger threshold, i.e., the largest impact parameter that still results in merger for each boost factor $\gamma$. The blue line shows the impact parameter for maximal radiation. The shaded region for both cases illustrates the sampling error in impact parameter. The two curves correspond exactly to those in the lower panel of Fig.~5 in the main text. This helps to better show the functional form of the maximal radiation and merger threshold branches.}
  \label{fig:map}
\end{figure*}

As emphasized in the main text, the phenomenology of ultra-relativistic encounters is characterized by a distinct separation between the prompt merger threshold and the impact parameter yielding maximal radiation efficiency for boosts $\gamma \gtrsim 3$. Fig.~\ref{fig:map} provides a 2D parameter space map of the radiation efficiency, clearly illustrating how the maximal radiation branch (blue line) diverges from the merger threshold (red dashed line) at higher Lorentz factors. 

\begin{figure}[t]
  \centering
  \includegraphics[width=0.5\textwidth]{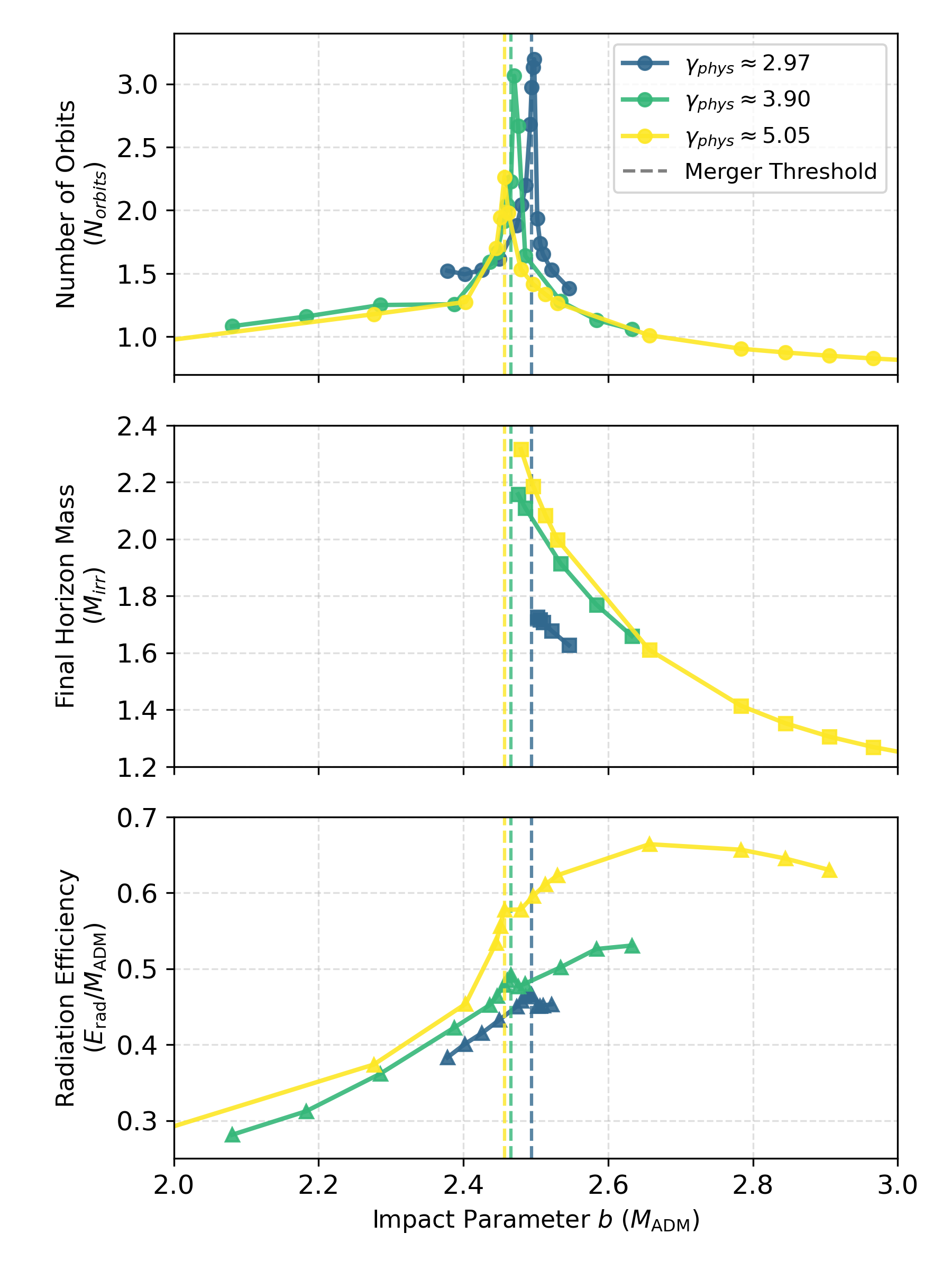}
  \caption{\textit{Top:} Number of orbits as a function of impact parameter; one can clearly see the sharp peak near the merger threshold, indicated by dashed lines. \textit{Middle:} Irreducible mass of the remnant black hole for the scattering cases, highlighting strong absorption at the merger threshold due to the increased number of interactions. 
  \textit{Bottom:} radiation efficiency as a function of impact parameter for the three selected boost values (gray dashed lines in Fig.~\ref{fig:map}). One sees a local maximum at the merger threshold, however, the global maxima for the two higher-boost cases occur at larger impact parameters, due to the strong absorption at the merger threshold.}
  \label{fig:dynamics}
\end{figure}

To understand the physical mechanism driving this separation, Fig.~\ref{fig:dynamics} breaks down the scattering dynamics for three representative boosts. Near the merger threshold, the binary undergoes multiple close interactions (as indicated by the sharp peak in the number of orbits in the top panel). While this prolonged interaction generates substantial gravitational radiation, the multi-pass lensing effect inside the transient trapping region also leads to significant horizon absorption (middle panel). Consequently, the net radiated energy escaping to infinity is suppressed near the threshold. The global maximum for radiation efficiency is thus pushed to slightly larger impact parameters (bottom panel), where the system still interacts violently enough to radiate copious amounts of energy, but avoids the strong horizon absorption associated with the near-threshold zoom-whirl behavior.

\section{Spin measurement}

To characterize the end state of the black holes discussed in Fig.~4 of the main text, we include the formula for computing spin through the quasi-local angular momentum:
\begin{equation}\label{spin_local}
    J[\varphi] = \frac{1}{8\pi}\oint_{\mathcal{H}} K_{ij} \varphi^j n^i dA~,
\end{equation}
where $\varphi^i$ is the approximate rotational Killing vector and $n^i$ is the outward-pointing unit normal to the horizon two-surface within the spatial slice, and through the geometric shape of the horizon applying a formula valid for stationary Kerr black holes:
\begin{equation}\label{spin_ce}
   J=2M_{\rm irr}\left(\frac{C_e}{4\pi}\right)\sqrt{\left(\frac{C_e}{4\pi}\right)^2-M_{\rm irr}^2}~,
\end{equation}
where $C_e$ is the equatorial proper circumference of the apparent horizon.

\section{Numerical Setup and Convergence}
To evolve the binary initial data, we use $6^{\rm th}$-order accurate finite differencing to evaluate spatial derivatives, and RK4 to integrate in time. 
For the production runs, the base grid extends to $\pm 1024~M_0$, with 216 grid points in each dimension;
with 9 levels of adaptive mesh refinement, we achieve a resolution of $\frac{1}{27}M_0$, or roughly $\frac{1}{275}M_{\rm ADM}$ for the set of simulations with highest boost ($\gamma\approx 5.1$).

To demonstrate convergence, we perform simulations with three different root-grid resolutions, decreasing to 204 and 192 points per dimension for the medium- and low-resolution runs, respectively, with identical refinement criteria and therefore the same mesh structure. 

\begin{figure}[t]
  \centering
  \includegraphics[width=0.5\textwidth]{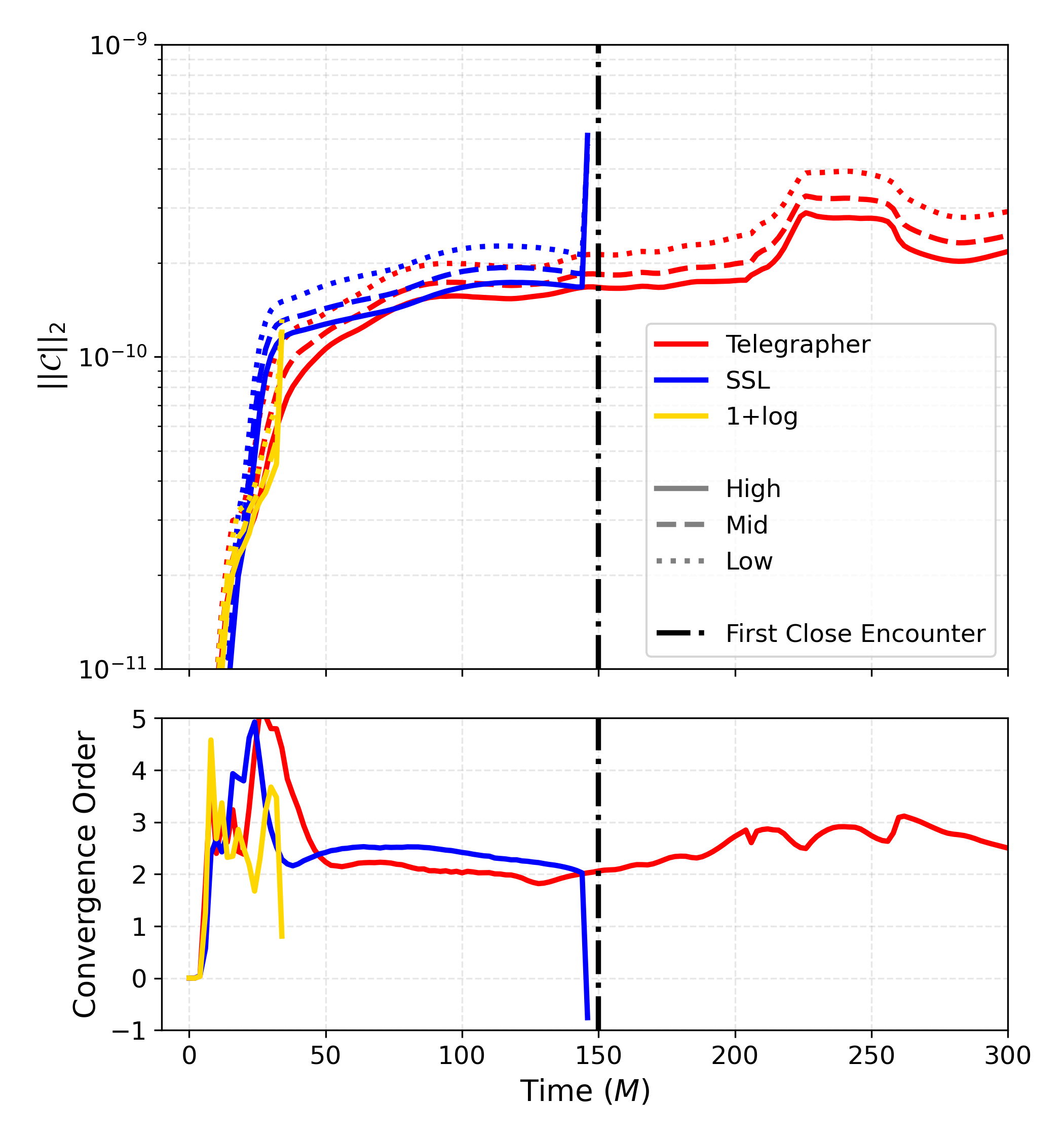}
  \caption{$L^2$ norm of the Z4c collective constraint (top) and the estimated convergence order (bottom) for the three gauge conditions at $\gamma\approx 5.1$. While the telegrapher gauge maintains stability and convergence throughout, the $1+\log$ and SSL crash at different stages of evolution. At very early times, the evolution-generated truncation error is subdominant to the constraint violation inherited directly from the initial data solver. Because the initial data is generated at the same fixed resolution across all three evolution resolutions, the apparent convergence order evaluates to zero near $t=0$. 
  The evolution-generated constraint then becomes more dominant after 20 $M_0$ of evolution, and the constraints return to high-order convergence.}
  \label{fig:conv}
\end{figure}

As highlighted in the main text, a major technical obstacle in simulating ultra-relativistic encounters is the emergence of severe gauge pathologies when highly boosted punctures pass through each other's strong-field regions. 
Fig.~\ref{fig:conv} demonstrates the necessity of the telegrapher gauge in this regime by plotting the constraint violation as a function of evolution time for a representative ``extreme'' case: a merger with $\gamma\approx 5.1$ and impact parameter $b=2.45 M_{\rm ADM}$, just below the merger threshold. 
The collective constraint for Z4c is defined as $\mathcal{C} \equiv \sqrt{\mathcal{H}^2+\mathcal{M}^2+\Theta^2 + \mathcal{Z}^2}$, see, e.g., Eqs.~18--21 of Ref.~\cite{Daszuta:2021ecf}. The $L^2$-norm is calculated as the volume integral of the square over the entire simulation domain, with the vicinity of the puncture excised, and normalized by the proper volume at each time. 

We see that while the new telegrapher gauge maintains stability and convergence throughout the violent near-zone dynamics, the ``vanilla'' $1+\log$ and slow start lapse (SSL) crash at different stages of the evolution. 
The $1+\log$ slicing does not survive the initial gauge dynamics, crashing after less than $30~M_0$ of evolution. 
While the SSL attenuates the initial gauge transformation through the inclusion of a damping term, the damping term is reduced over time, and the slicing condition converges back to $1+\log$.
Subsequently, the emergent strong gauge dynamics at the first close encounter cause the code to crash consistent with the development of a coordinate pathology. 

\begin{figure*}[t]
  \centering
  \includegraphics[width=1\textwidth]{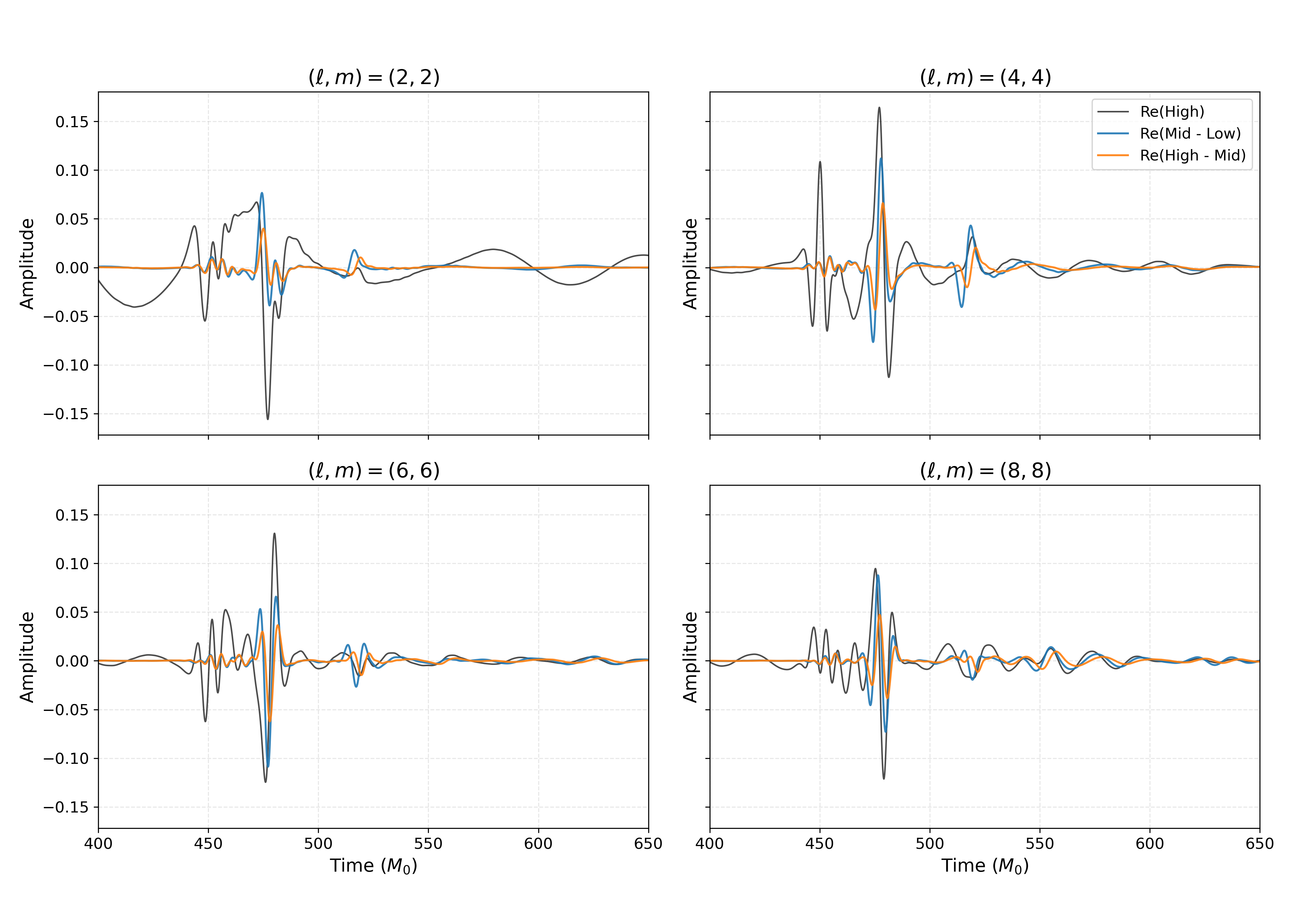}
  \caption{Convergence of dominant waveform modes with the telegrapher gauge for the $\gamma\approx 5.1$ prompt merger encounter. The black lines show the real part of the waveform for the highest resolution run, while the blue and orange lines show the differences between the medium- and low-resolution waveforms, and between the high- and medium-resolution waveforms, respectively. For this test, all three cases utilize an identical, fixed mesh refinement structure, with the base grid resolutions scaled proportionately. While the global constraint violation shown in Fig.~\ref{fig:conv} drops to lower order due to strong-field gradients and refinement boundaries, the waveforms extracted in the smooth wave-zone exhibit the expected 6th order convergence of the spatial differencing scheme. Furthermore, because relative errors can appear large during amplitude minima, estimating the relative error of the highest resolution case as a function of time provides a more robust measure of waveform accuracy. 
  While the relative waveform errors can still appear large, we show in Fig.~\ref{fig:rad_conv} that the uncertainty in the extracted radiated energy and angular momentum are both at the percent level.}
  \label{fig:conv_wf}
\end{figure*}

Fig.~\ref{fig:conv_wf} further confirms that the physical observables extracted in the wave zone, such as the highly irregular gravitational waveforms shown in Fig.~1 of the main text, are numerically robust. 
While the global constraint violation in Fig.~\ref{fig:conv} exhibits lower-order convergence (roughly 3rd order), the waveforms themselves exhibit approximately $6^{\rm th}$-order convergence of the underlying spatial differencing scheme.

\begin{figure*}[t]
  \centering
  \includegraphics[width=1\textwidth]{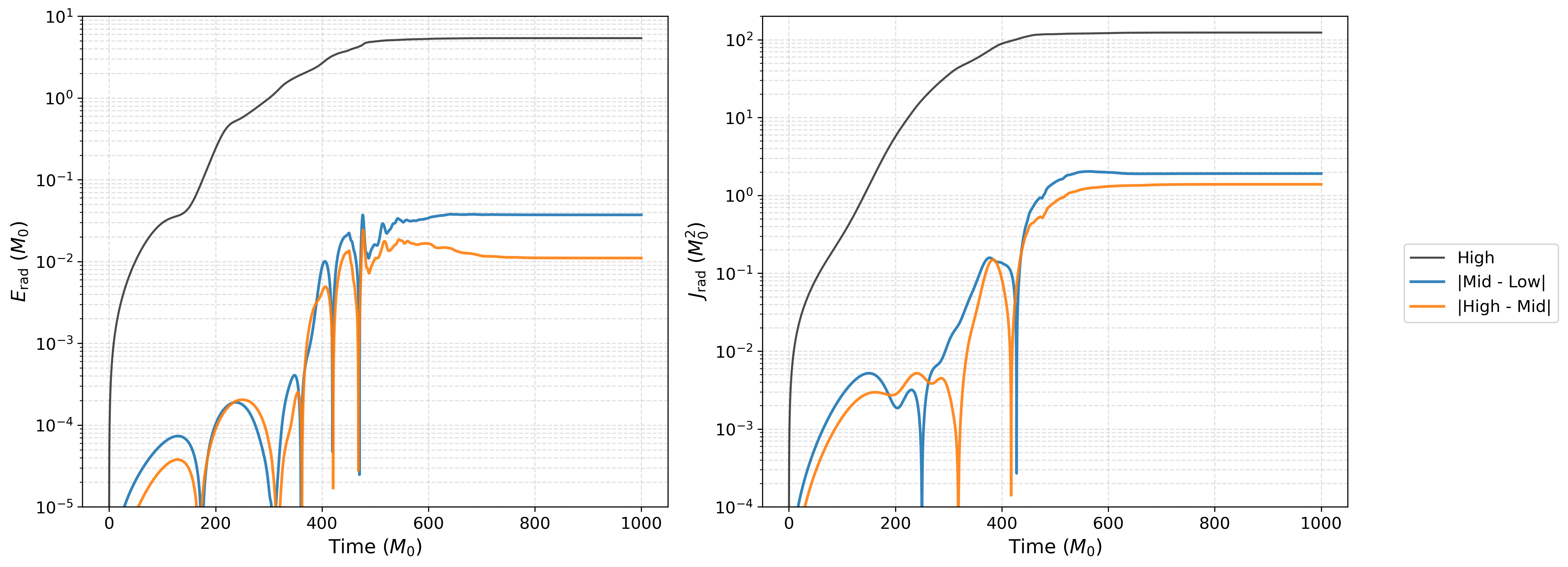}
  \caption{Accumulated radiated energy (left) and angular momentum (right) at three different resolutions. 
  The truncation error from the evolution is on the order of 1\%, much less than that associated with the junk radiation subtraction despite the highly irregular waveform. }
  \label{fig:rad_conv}
\end{figure*}

Lastly, in Fig.~\ref{fig:rad_conv}, we show that the error in the radiated energy and angular momentum due to finite grid resolution is less than 1\%, much smaller than the contamination due to the junk radiation. 

\section{Additional videos}

A video version of the super-Poynting-flux dynamics, i.e., Fig.~\ref{fig:superpoynting}, can be found \href{https://www.youtube.com/watch?v=6KtMbe4b9sc}{here}, and a volume rendering of the same field can be found \href{https://www.youtube.com/watch?v=B0ySMwM0HbA}{here}. 
The evolution of null integral curves that illustrates the formation of a transient light-ring region and strong trapping for the same run can be found \href{https://www.youtube.com/watch?v=-fmlEsOpj6U}{here}. 
A comparatively mildly relativistic run can be found \href{https://www.youtube.com/watch?v=AK5vuGw3P7U}{here}, where the trapping is significantly less strong, although the black holes have the same pre-interaction irreducible mass in both cases.

\bibliographystyle{apsrev4-2}
\bibliography{thebib}

\end{document}